\documentclass[preprint,review,3p,12pt]{elsarticle}

\usepackage{amssymb}

\journal{Journal of Crystal Growth}

\begin{document}

\begin{frontmatter}



\title{Single crystal growth of the hexagonal manganites $R$MnO$_3$
($R$ = rare earth) by the optical floating-zone method}


\author{C. Fan}
\author{Z. Y. Zhao}
\author{J. D. Song}
\author{J. C. Wu}
\author{F. B. Zhang}
\author{X. F. Sun\corref{cor1}}


\address{Hefei National Laboratory for Physical Sciences at Microscale,
University of Science and Technology of China, Hefei, Anhui
230026, China}

\cortext[cor1]{Corresponding author. Tel.: 86-551-63600499, Fax:
86-551-63600499.\\ Email address: xfsun@ustc.edu.cn}


\date{\today}

\begin{abstract}

We report a study on the crystal growth of the hexagonal
manganites $R$MnO$_3$ ($R$ = Y, Lu, Ho, Er, and Tm) by using an
optical floating-zone method. It was found that high-quality
single crystals of $R$ = Y, Lu, and Ho could be easily grown with
essentially the same conditions as those reported in literature,
that is, with an atmosphere of normal pressure Ar and oxygen
mixture and a growth rate of 2--4 mm/h. However, these conditions
were not feasible for growing good crystals of $R$ = Er and Tm.
The chemical analysis indicated that it was due to an
off-stoichiometric phenomenon in the formed single crystals. We
used an effective and simple way to resolve this problem by
adjusting the nominal compositions of the polycrystal feed rods to
be 1--2\% rare-earth excess. The structures and physical
properties were characterized by X-ray diffraction, magnetic
susceptibility, specific heat, resistivity, and dielectric
constant measurements.

\end{abstract}


\end{frontmatter}



\newpage

\section{Introduction}

The hexagonal rare-earth manganites $R$MnO$_3$ ($R$ = Y, Lu, Ho,
Er, and Tm) with the space group of $P6_3cm$ have attracted much
interest in the past decades because of their multiferroicity
\cite{Van, Lottermoser, Schouchkov, Fiebig1, Salama, Yen, Meier,
Fiebig2, Wang1, Wang2, Schmid}. In these compounds, each Mn$^{3+}$
ion is surrounded by three in-plane and two apical oxygen ions (a
MnO$_5$ bipyramid) and then they form a triangular-lattice layer.
Two adjacent layers are separated by a rare-earth layer and the
$R^{3+}$ ions are located in two different crystallographic sites,
``2$a$" and ``4$b$" \cite{Van2, Yakel}. The $H - T$ phase diagrams
of $R$MnO$_3$ are rather complicated and are strongly dependent on
the magnetism of rare-earth ions \cite{Yen, Fiebig2, Yen2, Hur,
Lorenz, Brown}. Till now, there is still no complete understanding
of the low-temperature magnetic structures and their transitions
in $R$MnO$_3$, particularly in HoMnO$_3$ and ErMnO$_3$. For
further and deeper studies of the magnetic properties and the
mechanism of magnetoelectic coupling effect in $R$MnO$_3$,
high-quality single crystals are very important. The hexagonal
$R$MnO$_3$ single crystals studied in the previous works were
grown by using the flux method or the optical floating-zone
technique and the latter one seemed to produce better crystals
\cite{Yen, Sharma, Lim, Zhou, Kim, Katsufuji}. The crystal growth
using the floating-zone technique in these works were all carried
out in air at a rate of 2--4 mm/h. The typical size of the
obtained crystals was 30--50 mm long with a diameter of 5--6 mm.
However, it is not clear whether these conditions have already
been optimized.

In this work, we try to optimize the growth conditions of
hexagonal $R$MnO$_3$ single crystals ($R$ = Y, Lu, Ho, Er, and Tm)
by using the floating-zone method. It was found that very good
single crystals of YMnO$_3$, LuMnO$_3$, and HoMnO$_3$ can be
obtained by using the reported growth conditions. However, the
growths of ErMnO$_3$ and TmMnO$_3$ need some tricks, because the
formed single crystals are slightly rare-earth rich and the
crystal growths could not be successful with the compositions
$R$:Mn = 1:1 for raw rod. We finally obtained successful growths
by modifying the ratio of $R$:Mn in the feed rod. The structures
and physical properties of these crystals were characterized by
X-ray diffraction (XRD), magnetic susceptibility, specific heat,
resistivity, and dielectric constant measurements.

\section{Crystal Growth}

The polycrystal powders of hexagonal $R$MnO$_3$ ($R$ = Y, Lu, Ho,
Er, and Tm) were synthesized by the conventional solid-state
reaction. Stoichiometric mixtures of $R_2$O$_3$ (99.99$\%$) and
MnO$_2$ (99.99$\%$) powders were ground together and calcined in
air at 1050$^{\circ}$C for 24 hours, followed by two more times
grinding and calcining to make sure of getting single phase. Then
the $R$MnO$_3$ powders were pressed into $\sim$ 100 mm long rods
with diameter of $\sim$ 7 mm under 30 MPa hydrostatic pressure.
The raw rods were sintered at 1250$^{\circ}$C for 20 hours in air
and used as feed rod and seed rod. The crystal growth was carried
out in an optical floating-zone furnace with four 500W halogen
lamps and ellipsoidal mirrors (Crystal System Incorporation,
Japan).

We first tried to grow HoMnO$_3$ using the same conditions as
those in the literature \cite{Katsufuji}. The atmosphere for
growth was normal-pressure mixture of argon and oxygen with the
ratio of 4:1 and the growth rate was 3 mm /h. The feed and seed
rods were rotating in opposite directions at 10 and 15 rpm,
respectively. After starting growth for only 2 hours, the surface
of crystal rod became smooth and two large facets on the opposite
sides appeared. This indicated that large and nearly single domain
of the crystal could be easily formed in such a short time.
Finally, a several-millimeter-long single crystal with shining
surface, nice morphology, and long consecutive facets was
obtained, as shown in Fig. 1(c). The long facets on the crystal
bars were identified to be the hexagonal plane normal to the $c$
axis by using the X-ray Laue back diffraction. YMnO$_3$ and
LuMnO$_3$ single crystals with good appearances, as shown in Figs.
1(a) and 1(b), could also be obtained by using the same growth
conditions. After trying more, it was found that the growths of
these crystals were actually not very sensitive to the growth
rate; the rates of 2--4 mm/h gave similar products, which is
coincided with the literature. However, the crystal growths are
more sensitive to the atmosphere. Other atmospheres like pure
oxygen or pure argon were found to be not suitable. In pure
oxygen, for example, there were some bubbles in the molten zone
and the stable crystal growth was impossible. Small pieces of
these as-grown crystals were cut and ground into powders for XRD
and the data confirmed the pure and single phase. In addition, the
good crystallinity of these crystals were also confirmed by narrow
widths of the X-ray rocking curves. For example, the full width at
the half maximum (FWHM) of the rocking curve for the (004) Bragg
peak is about 0.10--0.14$^{\circ}$ for these crystals. (The XRD
data are not shown here but can refer to those of ErMnO$_3$, as
shown in Fig. 3.)

The main finding of this work is that the growths of ErMnO$_3$ and
TmMnO$_3$ single crystals are not so simple. It was found that
both compounds could not grow in a stable state by using the same
process as HoMnO$_3$, YMnO$_3$, and LuMnO$_3$. For example, Figs.
2(a-d) show the photographs which recorded the growth process of
ErMnO$_3$ with a rate of 2 mm/h. At the beginning, the growth
looked stable and is very similar to those of HoMnO$_3$, YMnO$_3$,
and LuMnO$_3$. After 8 hours, however, some folds appeared on the
feed rod. Then the feed rod started to expand with some unmelted
solids rolling up on the top edge of the molten zone, which is
indicated by a red arrow in Fig. 2(c). With the increase of the
unmelted solids, the molten zone became narrower and unstable. The
diameter of the single crystals also decreased at the same time,
as shown in Fig. 2(d). After 15 hours, the molten zone finally
dropped, which was apparently related to the change of viscosity.
In addition, small amount of gray powders were found to adhere on
the glass shield of the furnace; these were the volatile matters
during the growth. All these phenomena were not observable in the
growth of HoMnO$_3$, YMnO$_3$, and LuMnO$_3$ crystals. The
obtained ErMnO$_3$ crystal bar has many small cracks inside,
indicating a poor crystal quality. It was also found that changing
the growth rate or atmosphere could not be helpful for resolving
these problems.

We analyzed the chemical constituents of several important parts,
A, B, and C shown in Fig. 2(d), by using X-ray fluorescence
spectrometry. The measurements were done by using a Sequential
X-ray Fluorescence Spectrometer with Rh anode tube (XRF-1800,
Shimdzu). The excitation current is 90 mA and the analyzing
crystal is LiF$_{200}$. The area illuminated by the incident X-ray
beams is 20 mm in diameter. The data were calculated using
Fundamental Parameter (FP) method and then the proportion of the
rare-earth and Mn elements was obtained. Part A, the unmelted
solid on the edge of the feed rod, was found to have a molar ratio
of Er:Mn = 0.363:0.637, indicating that much rare-earth element
was lost in this part. The molar ratio of Er:Mn in part B was
0.497:0.503 rather than 0.5:0.5. The as-grown crystal, indicated
as part C in Fig. 2(d), had a molar ratio of Er:Mn = 0.508:0.492,
which is a bit different from the nominal composition of the raw
rod. In addition, the volatile matters on the glass shield were
found to be pure Mn oxide and have no Er element. This is
reasonable since the rare-earth oxides have boiling points above
2900 $^{\circ}$C and are difficult to be volatilized. According to
these results and considering the volatilization of Mn element, it
is clear that the feed rod of ErMnO$_3$ could not be melted
congruently. Instead, the formed ErMnO$_3$ crystal was about 0.8\%
Er rich. We therefore prepared new polycrystal powder with
slightly changing the stoichiometric proportion of Er:Mn to
0.5075:0.4925 and repeated the growth. This time the crystal
growth could be stable enough to finish the growth (up to 40
hours), as shown in Figs. 2(e-h). The as-grown ErMnO$_3$ single
crystal bar shows very nice appearance and with long consecutive
facets on the surface, as shown in Fig. 1(d). The powder XRD of
the as-grown crystal confirmed the pure and single phase, as shown
in Fig. 3(a). A narrow width of the rocking curve of (300) Bragg
peak (FWHM $\approx 0.13^\circ$), as shown in Fig. 3(b),
demonstrated that the ErMnO$_3$ single crystal has good
crystallinity.

The case of TmMnO$_3$ is very similar to ErMnO$_3$. Based on the
X-ray fluorescent spectrometry data, we adjusted the
stoichiometric proportion of Tm:Mn to 0.5085:0.4915 and nice
single crystals were also obtained, as shown in Fig. 1(e).

The off-stoichiometric growth of ErMnO$_3$ and TmMnO$_3$ single
crystals has never been reported in previous literature. However,
it is not rare in binary-element systems. As an example, a similar
case is growing LiNbO$_3$ crystal from Li$_2$O-Nb$_2$O$_3$ system,
in which the congruent melting is located at the molar composition
of Li:Nb = 0.486:0.514 \cite{Lemer, Svaasand}. In this case, it
crystallizes with an off-stoichiometry of
Li$_{0.945}$NbO$_{2.973}$. Another example is the
La$_2$O$_3$-Mn$_2$O$_3$ system for growing orthorhombic LaMnO$_3$
\cite{Roosmalen}. Apparently, in the phase diagrams of the
hexagonal ErMnO$_3$ and TmMnO$_3$, the congruent melting points
are located at the rare-earth-rich side. Since there would be more
Er element crystallized into the single crystal, the percentage of
Er element decreased in the molten zone. Thus, the molten zone had
to absorb more Er element from the feed rod, inducing the
deficiency of Er element in the feed rod. This process could not
be continued for a long time when the molten zone was too lack of
Er. Modifying the compositions of the feed rods could effectively
resolve this problem. Our results give a more precise
understanding about the phase diagrams of hexagonal $R$-Mn-O
systems \cite{Balakirev}.

The optimized growth conditions for hexagonal $R$MnO$_3$ are
summarized in Table 1. The lattice parameters determined by the
powder XRD are also shown in the table and are in good
correspondences with the results in previous works. Our crystals
have not only good appearances but also good crystallinity, as the
XRD and Laue photographs indicated.

\section{Magnetic susceptibility and specific heat}

The basic physical properties of the obtained hexagonal $R$MnO$_3$
single crystals were also characterized. DC magnetization and
specific heat measurements were done by using a SQUID-VSM (Quantum
Design) and a Physical Property Measurement System (PPMS, Quantum
Design), respectively. Resistivity was measured by the standard
four-probe technique. Dielectric constants were measured by using
a precision impedance analyzer (Agilent 4294A).

Fig. 4 shows the temperature dependencies of the magnetic
susceptibilities along the $a$ and $c$ axes of YMnO$_3$,
HoMnO$_3$, ErMnO$_3$, and TmMnO$_3$. YMnO$_3$ shows much smaller
susceptibility than other materials, which means that the
rare-earth ions make main contributions in $R$MnO$_3$. The
$\chi_a(T)$ and $\chi_c(T)$ of YMnO$_3$ show clear transitions at
72 K, which is due to the antiferromagnetic (AF) transition of
Mn$^{3+}$ moments \cite{Fiebig1, Muoz}. For HoMnO$_3$, ErMnO$_3$,
and TmMnO$_3$, the AF transitions of Mn$^{3+}$ moments at about
70--90 K can hardly be distinguished from the $\chi(T)$ data, as
shown in Figs. 4(b-d). One noticeable transition in the data of
these crystals is a weak slope change of $\chi_c(T)$ of HoMnO$_3$
at $\sim$ 40.5 K, which is due to a spin re-orientation of
Mn$^{3+}$ sublattice from $P6'_3c'm$ to $P6'_3cm'$. The Ho$^{3+}$
moments form a long-range AF order at about 4.2 K, where the
slopes of $\chi_a(T)$ and $\chi_c(T)$ show some rather clear
changes. In ErMnO$_3$, the $\chi_c$ shows a quick increase below
about 10 K. However, there is no clear indication on $\chi(T)$
curves whether a long-range order is established at temperatures
down to 2 K. The Tm$^{3+}$ moments do not show long-range order at
temperatures down to 2 K. However, the low-$T$ susceptibilities
show deviations from the paramagnetic behavior below 10 K. These
suggest that the magnetic correlations among Tm$^{3+}$ moments are
not negligible.

Fig. 5 shows the magnetization curves of YMnO$_3$, HoMnO$_3$,
ErMnO$_3$, and TmMnO$_3$ at 2 K with the magnetic field applied
along the $a$ and $c$ axes. The $M(H)$ curves of YMnO$_3$ show no
field-induced transition but only a linear behavior up to 7 T.
Moreover, the magnetization at 7 T is very small, which is far
from the theoretical value (4 $\mu_B$) of free Mn$^{3+}$ ions.
This is because the Mn$^{3+}$ ions have formed a 120$^\circ$ AF
arrangement in the $ab$ plane at 72 K \cite{Fiebig1, Muoz}, and 7
T is too low to polarize the Mn$^{3+}$ moments. The behaviors for
the other three compounds with magnetic Ho$^{3+}$, Tm$^{3+}$, and
Er$^{3+}$ ions are more complicated.

For TmMnO$_3$, the magnetization curves show strong anisotropy.
The magnetization in the $a$-axis field is much smaller than that
in the $c$-axis field, which is supportive to a strong
Ising-anisotropy of the Tm$^{3+}$ on 4$b$ sites \cite{Yen,
Skumryev}. The $M(H)$ curve along the $a$ axis is close to a
linear function up to 7 T except for a small curvature at low
fields, which is mainly the paramagnetic response of the 2$a$
moments. In contrast, a clear transition is shown at 3.5 T in the
$M(H)$ curve for $H \parallel c$, which is coincided with a
field-induced transition of Mn$^{3+}$ magnetic structure from
$P6'_3c'm$ to $P6_3c'm'$ \cite{Yen}. This is directly related to a
spin-flop transition of the 4$b$ moments. The slower increase of
magnetization above this field is likely the paramagnetic
contribution from the 2$a$ moments.

For ErMnO$_3$, the earlier studies suggested that at low
temperatures Er$^{3+}$ moments at the 2$a$ and 4$b$ sites could
form respective long-range AF order with the easy axis along the
$c$ direction \cite{Yen}. As shown in Fig. 5(c), the magnetization
for $H \parallel a$ is much larger than that for $H \parallel c$,
and it just shows a simple spin polarization process up to 7 T
with the magnitude of 7.4 $\mu_B$ per formula. This seems to be in
contradictory with the proposed Ising anisotropy of Er$^{3+}$
moments. The $M(H)$ curve for $H \parallel c$ is consistent with
the data in previous literature \cite{Meier} and shows two
transitions at 0.01 and 0.85 T, which are related to the magnetic
transitions from the so-called FIM$_1$ state (multi-domain of
$\Gamma \rm_2^{Er}$) to FIM$_2$ state (single domain of $\Gamma
\rm_2^{Er}$) then to the ferromagnetic state.

The case of HoMnO$_3$ is somewhat similar to ErMnO$_3$ in the
regard that the magnetization for $H \parallel a$ is much larger
than that for $H \parallel c$ and it shows a simple spin
polarization process with the magnitude of 7.8 $\mu_B$ per formula
at 7 T. This is also in contradictory with the proposed Ising
anisotropy of Ho$^{3+}$ moments. The magnetization curve is more
complicated in the $c$-axis field. It shows two step-like
transitions at 0.85 and 2.5 T, respectively, which are in good
correspondence with the field-induced transitions proposed from
the earlier dielectric and heat transport measurements \cite{Yen,
Wang2, Fiebig3}. If Ho$^{3+}$ ions have weak spin anisotropy,
these two transitions could be spin flops of Ho$^{3+}$ moments at
the 2$a$ and 4$b$ sites, associated with the spin re-orientations
of Mn$^{3+}$ spin structures. There would be another transition at
higher field above 7 T, which is due to the polarization of
Ho$^{3+}$ spins \cite{Yen, Wang2}. Actually, Kim et al. have
indeed observed a third transition of $M(H)$ curve at about 6.5 T
in HoMnO$_3$ thin film. The lower transition field might be due to
the difference between bulk and film samples \cite{Kim2}.

Low-temperature specific heat data of $R$MnO$_3$ ($R$ = Y, Lu, Ho,
Er, and Tm) crystals are shown in Fig. 6. They all show a
transition at the temperatures ranging from 69 to 87 K, which are
attributed to the AF order of Mn$^{3+}$ moments. Below 20 K, as
shown in the inset to Fig. 6, there is no transition down to 2 K
for YMnO$_3$ and LuMnO$_3$, since both Y$^{3+}$ and Lu$^{3+}$ are
nonmagnetic. For TmMnO$_3$, there is a hump-like feature at about
4--16 K, which is related to the development of short-range
correlation of Tm$^{3+}$ moments \cite{Wang1}. HoMnO$_3$ shows a
peak at about 4.5 K, indicating the long-range AF order of
Ho$^{3+}$ moments. For ErMnO$_3$, there are a small peak at 2 K
and a broader peak at 3.2 K, which indicate the transition from
the AF state to the FIM$_1$ state and the short-range correlation
of Er$^{3+}$ moments at the 2$a$ sites, respectively \cite{Meier}.

Fig. 7 shows the temperature dependencies of the $c$-axis
resistivity of $R$MnO$_3$ ($R$ = Y, Ho, Er, and Tm) single
crystals. It can be seen that the resistivity values of these
manganites are very large and increase with lowering temperature
roughly in a exponential way, indicating good insulating behaviors
\cite{Katsufuji}.

Low-temperature dielectric constants along the $c$ axis and at 50
kHz of $R$MnO$_3$ ($R$ = Y, Ho, Er, and Tm) single crystals are
shown in Fig. 8. The main features are the slope changes of data
at about 70--80 K and the lower-temperature cusps. It has been
known that they are related to the N\'eel transitions of Mn$^{3+}$
moments and some spin re-orientations of Mn$^{3+}$ions,
respectively \cite{Yen, Katsufuji, Ghosh, Sahu, Sugie, Iwata}.

Apparently, these magnetic, specific heat, and electric data are
essentially in good agreement with most of the previous reports.
This means that our single crystals are comparable to those from
other groups. One may note that the off-stoichiometric
compositions of ErMnO$_3$ and TmMnO$_3$ in principle should affect
the low-temperature physical properties, since about 1\%
rare-earth ions occupy the Mn$^{3+}$ sites. In the common
measurements, such effect seems to be not yet detected. It calls
for further investigations on the relationship between the
microstructure and the magnetic properties of the hexagonal
rare-earth manganites.

\section{CONCLUSIONS}

High-quality single crystals of the hexagonal $R$MnO$_3$ ($R$ = Y,
Lu, Ho, Er, and Tm) were grown by the optical floating-zone
method. YMnO$_3$, LuMnO$_3$, and HoMnO$_3$ single crystals could
be grown very easily. However, ErMnO$_3$ and TmMnO$_3$ single
crystals showed some difficulty in growth process. Chemical
analysis indicated that it was due to the off-stoichiometric
compositions of these two single crystals. The problem was
resolved by modifying the nominal compositions of the polycrystal
feed rod. The structure and crystallinity of obtained crystals
were characterized by X-ray diffraction and Laue photographs. The
low-temperature magnetization, specific heat, resistivity, and
dielectric constant were measured to characterize the basic
physical properties.

\section*{ACKNOWLEDGMENTS}

This work was supported by the National Natural Science Foundation
of China, the National Basic Research Program of China (Grant Nos.
2009CB929502 and 2011CBA00111), and the Fundamental Research Funds
for the Central Universities (Program No. WK2340000035).






\bibliographystyle{elsarticle-num}
\bibliography{<your-bib-database>}





\newpage

\begin{figure}[tp]
\vglue 1.0cm
\newpage


\caption{(Color online) Photographs of $R$MnO$_3$ ($R$ = Y, Lu,
Ho, Er, and Tm) single crystals.}


\caption{(Color online) The photographs recording the growth
process of ErMnO$_3$ single crystals. (a-d) The composition of the
feed rod is ErMnO$_3$ and growth rate is 2 mm/h. The red arrows
indicate three different areas A, B, and C, which are the unmelted
solids on the feed rod, the molten zone, and the crystal,
respectively. The molten zone was finally dropped after about 15
hours. (e-h) The composition of the feed rod is
Er$_{1.015}$Mn$_{0.985}$O$_3$. The growth rate is also 2 mm/h. The
growth was very stable up to 40 hours when it was finished.}

\caption{(a) X-ray diffraction of powders ground from a piece of
ErMnO$_3$ single crystal. (b-c) X-ray diffraction pattern of
($h$00) plane and the rocking curve of (300) peak.}

\caption{(Color online) Magnetic susceptibilities of (a) YMnO$_3$,
(b) HoMnO$_3$, (c) ErMnO$_3$, and (d) TmMnO$_3$ single crystals
measured with 0.1 T magnetic field applied along the $a$ or $c$
axis. The arrows indicate the weak changes in the slope of
$\chi(T)$ curves. Among them, the slope change in panel (b) is due
to a spin-reorientation of Mn$^{3+}$ sublattice ($T_{SR}$), while
those in other panels are due to the AF transition of Mn$^{3+}$
moments ($T_N$).}

\caption{(Color online) Magnetization curves of (a) YMnO$_3$, (b)
HoMnO$_3$, (c) ErMnO$_3$, and (d) TmMnO$_3$ single crystals at 2 K
with the magnetic field applied along the $a$ or $c$ axis. }

\caption{(Color online) Temperature dependencies of specific heat
of $R$MnO$_3$ ($R$ = Y, Lu, Ho, Er, and Tm) single crystals.}

\caption{(Color online) Temperature dependencies of the $c$-axis
resistivity of $R$MnO$_3$ ($R$ = Y, Ho, Er, and Tm) single
crystals.}

\caption{(Color online) Temperature dependencies of the dielectric
constant along the $c$ axis and at 50 kHz of $R$MnO$_3$ ($R$ = Y,
Ho, Er, and Tm) single crystals. The arrows show the slope changes
of data or lower-temperature cusps; the former ones are related to
the N\'eel transitions of Mn$^{3+}$ moments while the latter ones
are related to some spin re-orientations of Mn$^{3+}$ions. }

\end{figure}
\clearpage

\begin{figure*}[htbp]
\center {$\Huge\textbf{Fig. 1} $}

\includegraphics[bb = 10 800 700 500, width=1.2\textwidth]{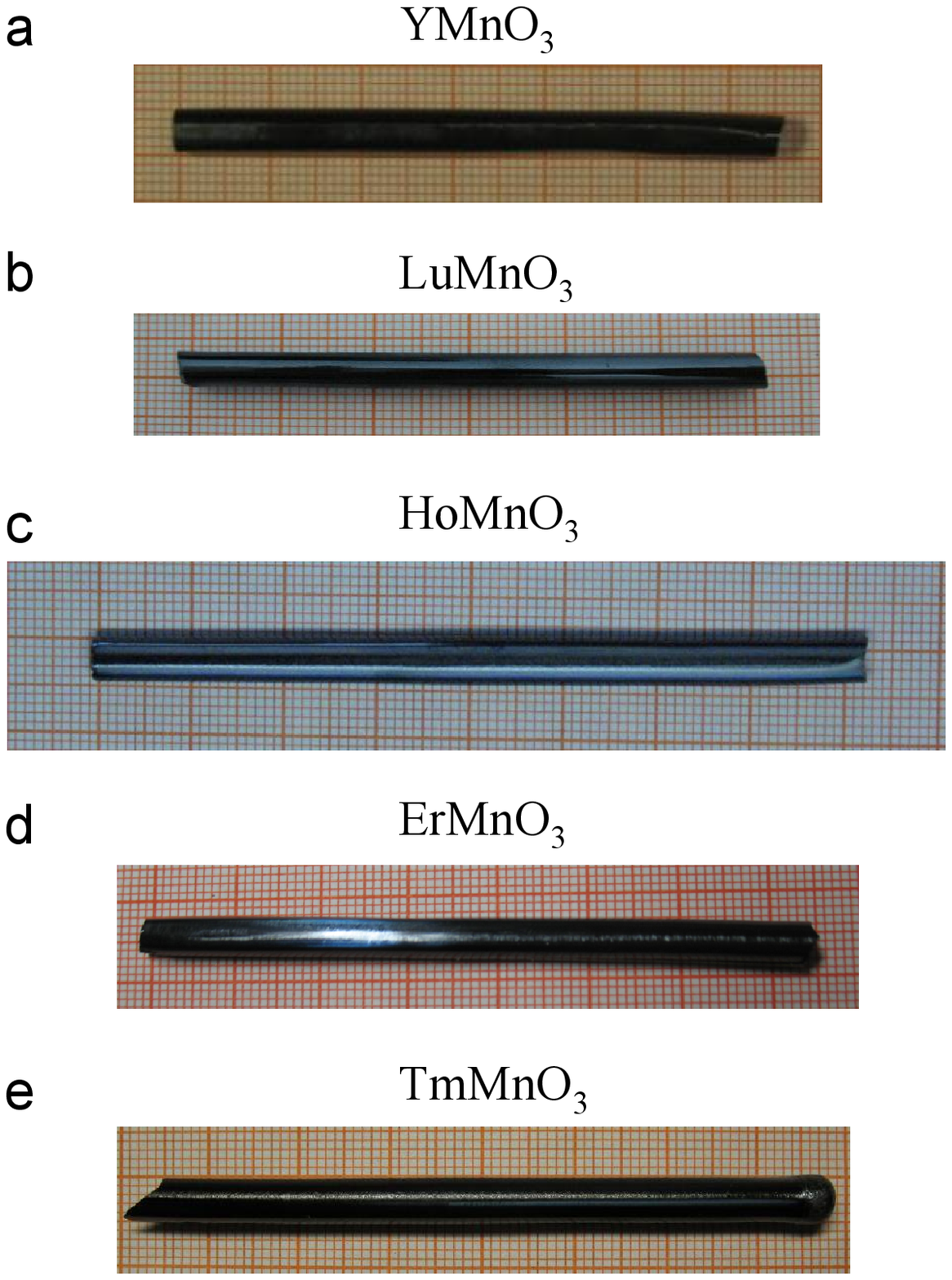}
\end{figure*}
\clearpage
\newpage
\begin{figure*}[htbp]
\center {$\Huge\textbf{Fig. 2} $}

\includegraphics[bb = 10 800 700 500, width=1.2\textwidth]{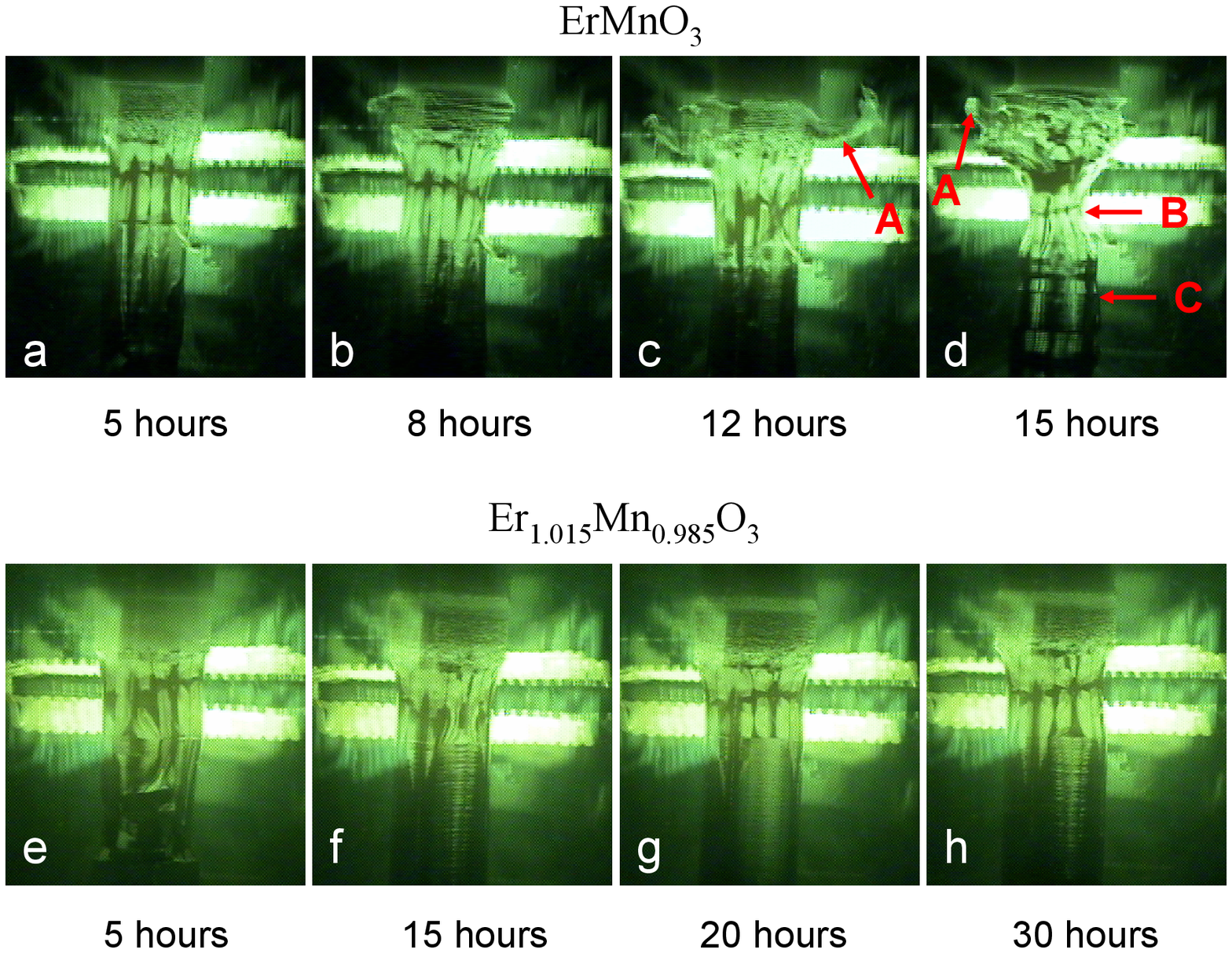}
\end{figure*}
\clearpage
\newpage
\begin{figure*}[htbp]
\center {$\Huge\textbf{Fig. 3} $}

\includegraphics[bb = 10 800 700 700, width=1.0\textwidth]{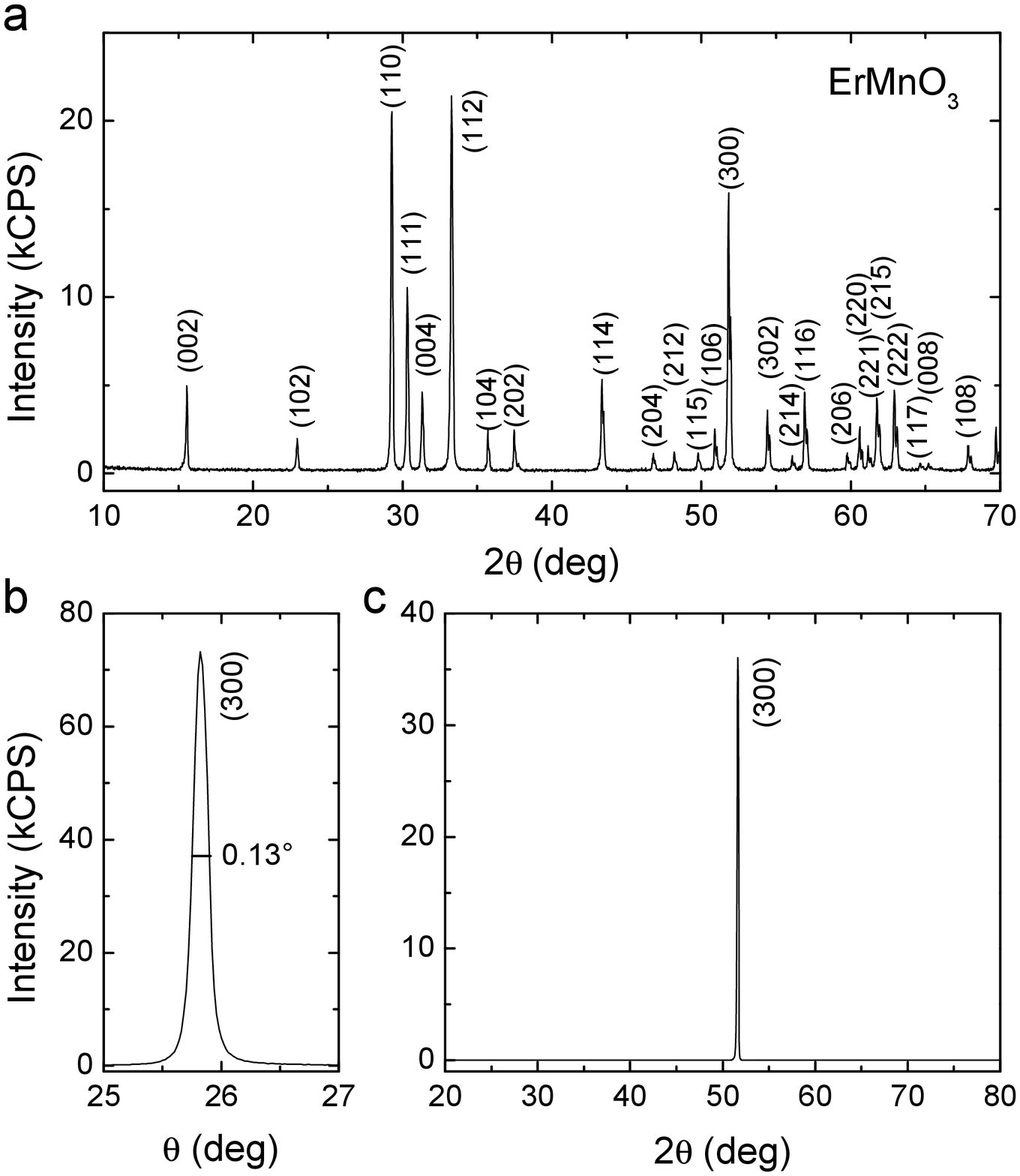}
\end{figure*}
\clearpage
\newpage
\clearpage
\newpage
\begin{table}[htbp]
\caption {Optimized growth conditions for different rare-earth
hexagonal manganites. The lattice parameters are obtained from the powder X-ray diffraction of single crystals.} \centering
\begin{tabular}{p{60pt}p{50pt}p{50pt}p{75pt}p{65pt}p{100pt}}
\hline & a ({\AA}) \centering & c ({\AA})\centering &
Stoichiometry ($R$:Mn) of feed rod \centering & Growth rate (mm/h)
\centering & Atomosphere\\\hline
  YMnO$_3$ & 6.139(4) \centering & 11.426(1) \centering & 1:1 \centering & 2--4 \centering & 1 atm
  Ar+O$_2$(4:1)\\
  HoMnO$_3$ & 6.138(6) \centering & 11.426(6) \centering & 1:1 \centering & 2--4 \centering & 1 atm Ar+O$_2$(4:1)\\
  ErMnO$_3$ & 6.106(4) \centering & 11.432(7) \centering & 1.015:0.985 \centering & 2 \centering & 1 atm Ar+O$_2$(4:1)\\
  TmMnO$_3$ & 6.082(7) \centering & 11.391(6) \centering & 1.017:0.983 \centering & 2 \centering & 1 atm Ar+O$_2$(4:1)\\
  LuMnO$_3$ & 6.040(8) \centering & 11.395(1) \centering & 1:1 \centering & 2--4 \centering & 1 atm Ar+O$_2$(4:1)\\
\hline \hline
\end{tabular}
\end{table}
\clearpage
\newpage
\begin{figure*}[htbp]
\center {$\Huge\textbf{Fig. 4} $}

\includegraphics[bb = 10 800 700 700, width=1.2\textwidth]{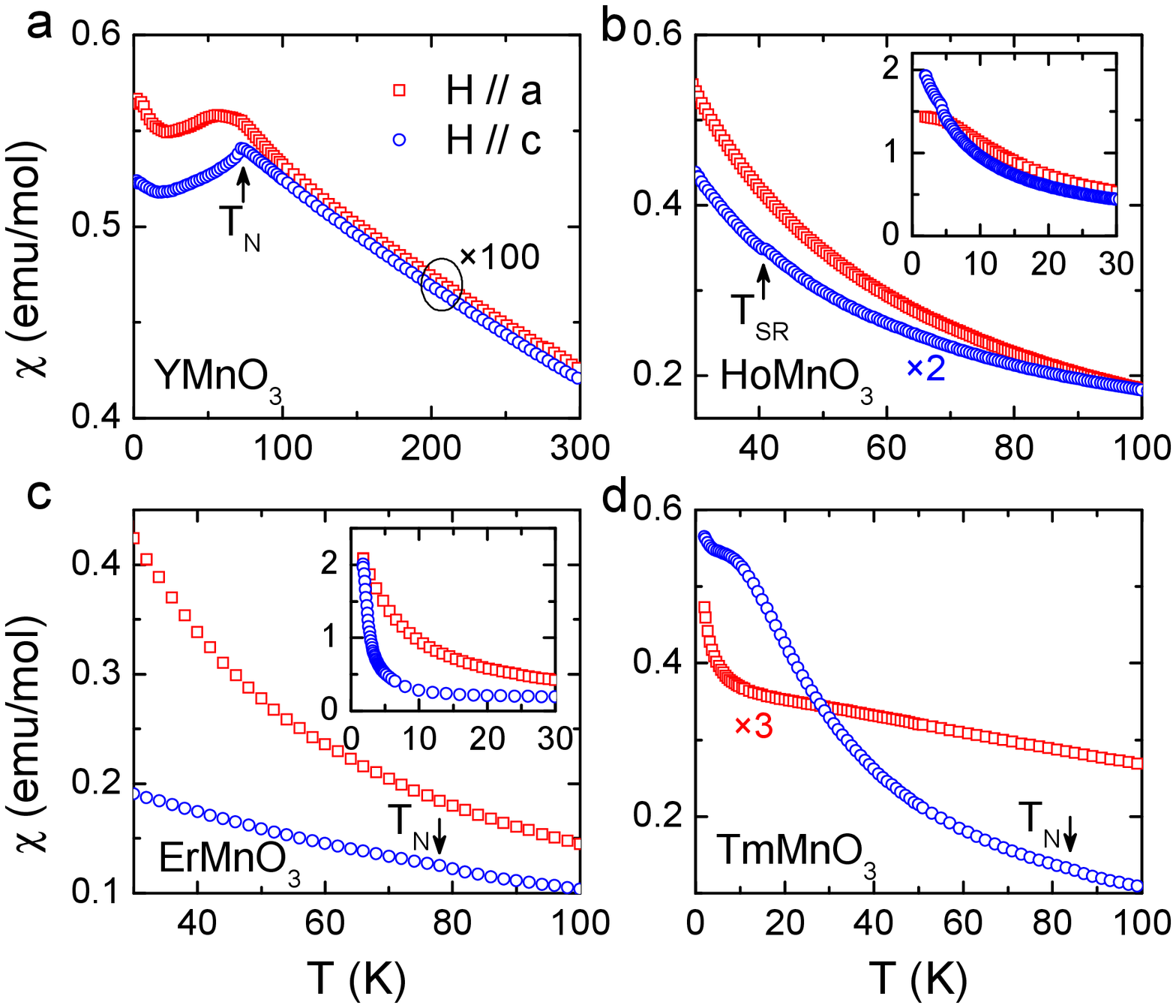}
\end{figure*}
\clearpage
\newpage
\begin{figure*}[htbp]
\center {$\Huge\textbf{Fig. 5} $}

\includegraphics[bb = 10 800 700 700, width=1.2\textwidth]{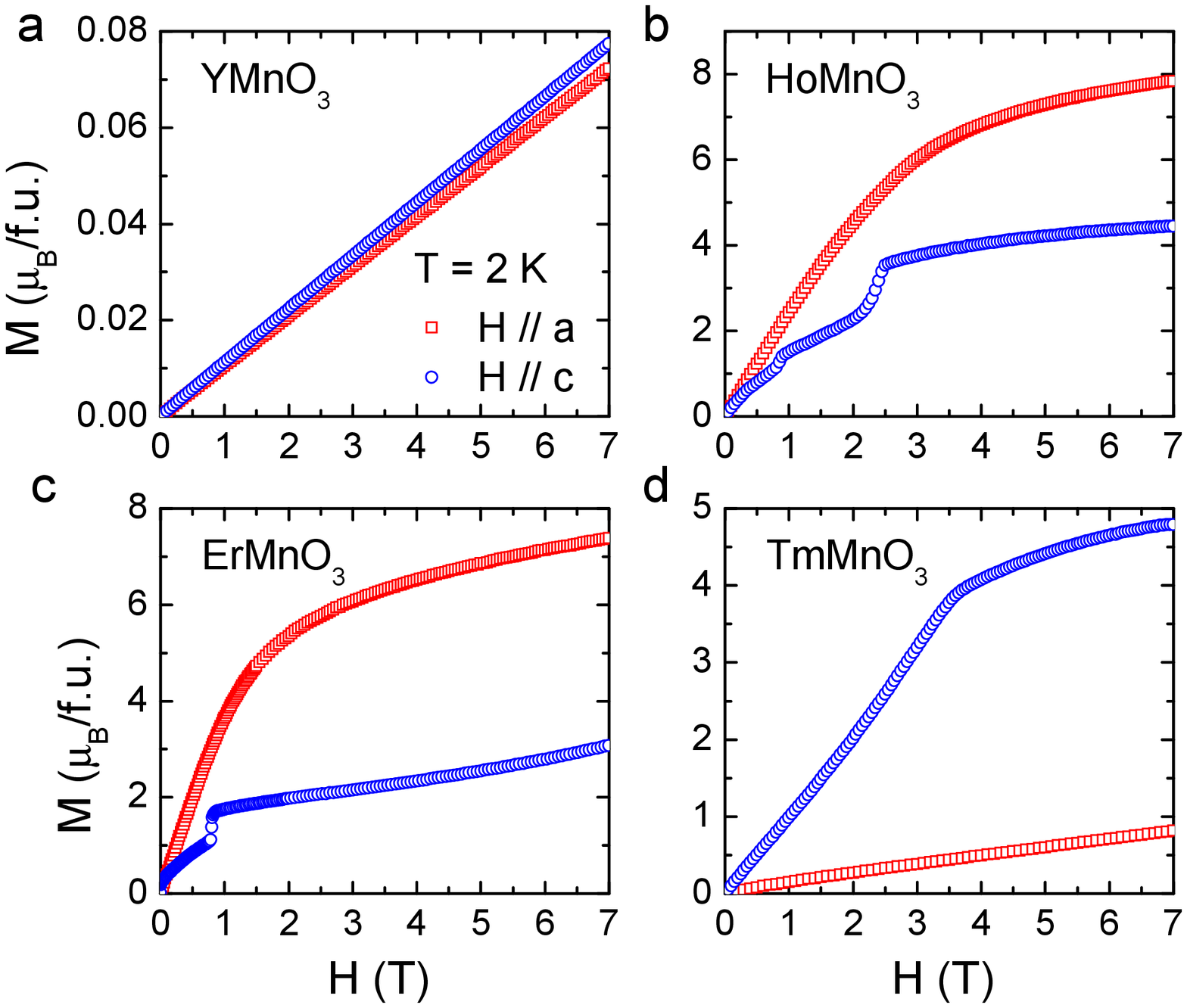}
\end{figure*}
\clearpage
\newpage

\begin{figure*}[htbp]
\center {$\Huge\textbf{Fig. 6} $}

\includegraphics[bb = 10 800 700 700, width=1.5\textwidth]{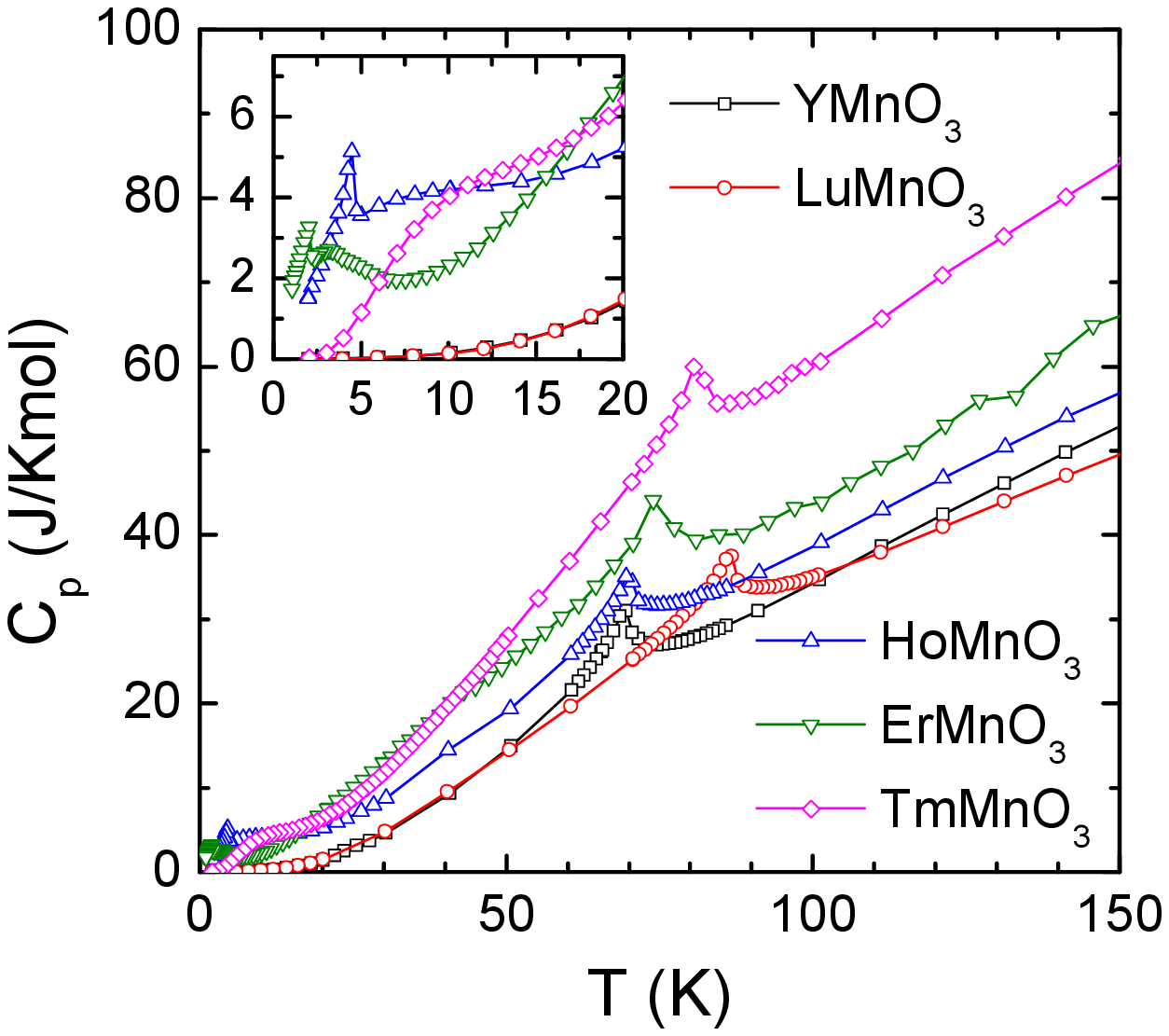}
\end{figure*}
\clearpage
\newpage

\begin{figure*}[htbp]
\center {$\Huge\textbf{Fig. 7} $}

\includegraphics[bb = 10 800 700 700, width=1.5\textwidth]{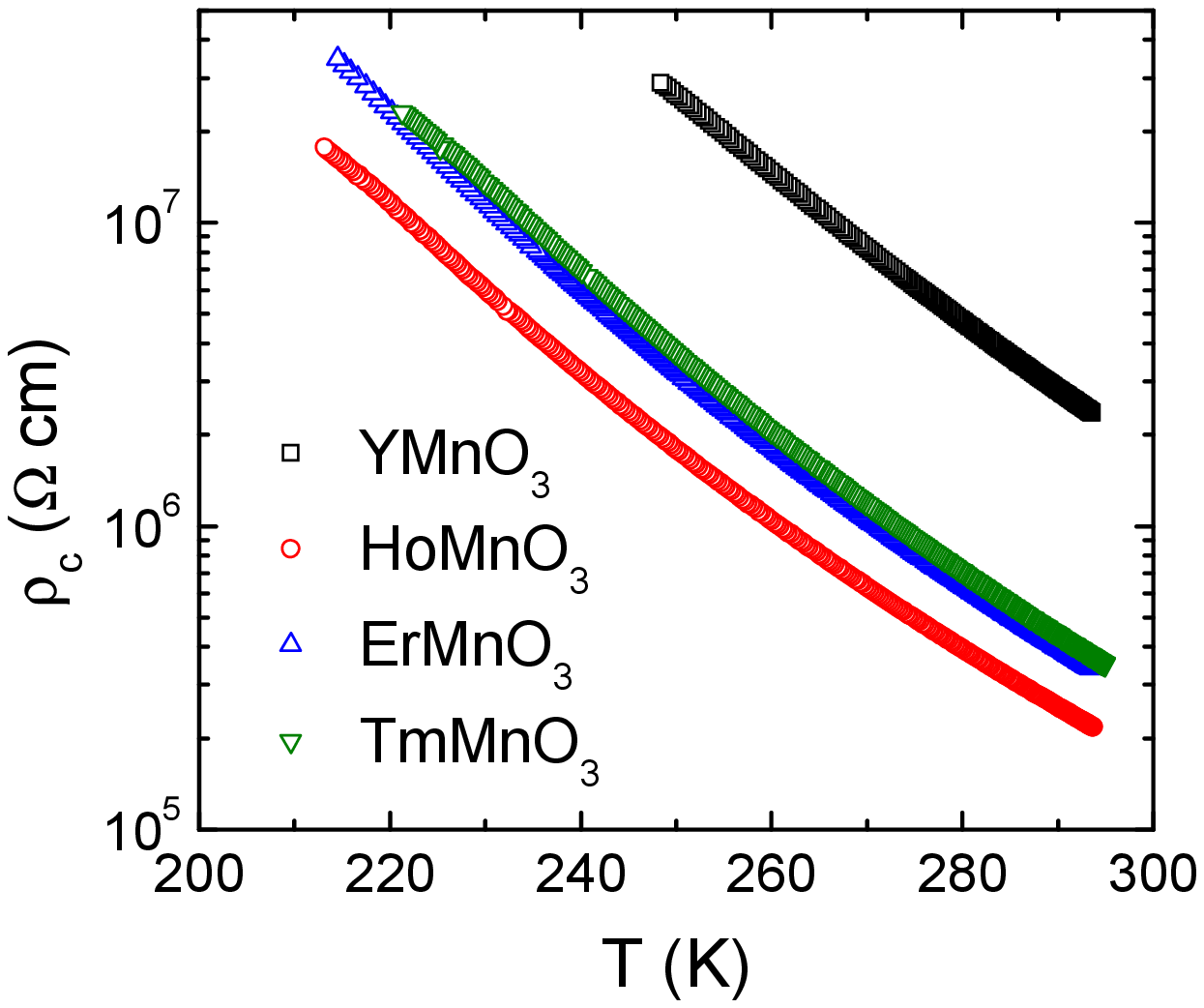}
\end{figure*}
\clearpage
\newpage

\begin{figure*}[htbp]
\center {$\Huge\textbf{Fig. 8} $}

\includegraphics[bb = 10 800 700 700, width=1.2\textwidth]{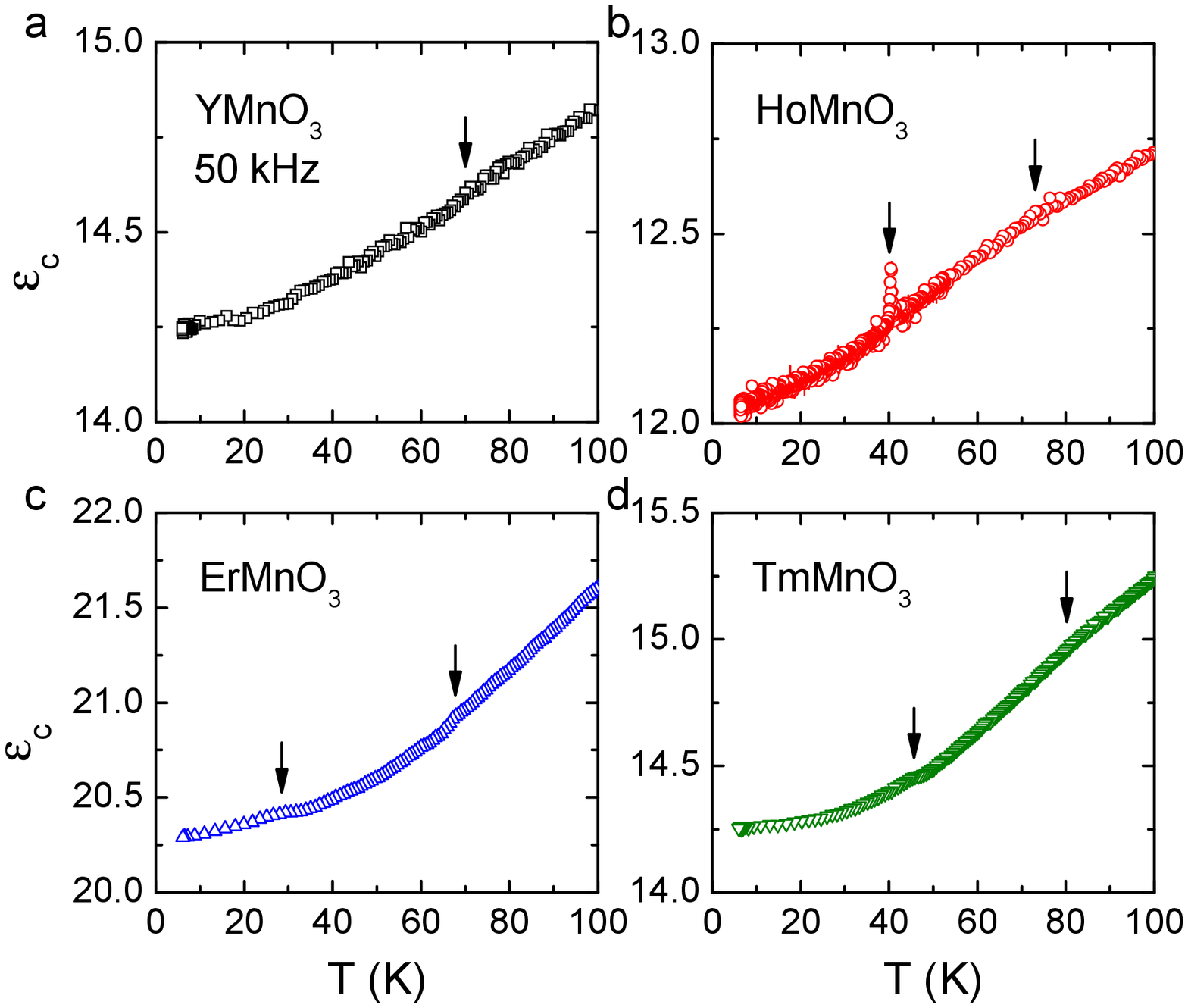}
\end{figure*}
\clearpage
\newpage

\end{document}